\documentclass[12pt]{book}
\usepackage{plenum}
\usepackage{epsf}
\begin{document}
\def\no{\nonumber}
\def\qu{\quad}
\def\qb{\bar{q}}
\def\qbm{\bar{\mbox{q}}}
\def\la{\langle}
\newcommand{\gm}{\gamma}
\newcommand{\be}{\begin{eqnarray}}
\newcommand{\ee}{\end{eqnarray}}
\renewcommand{\th}{\theta}
\newcommand{\Sg}{\Sigma}
\newcommand{\dl}{\delta}
\newcommand{\SSg}{\tilde{\Sigma}}
\newcommand{\eq}{\begin{equation}}
\newcommand{\eqx}{\end{equation}}
\newcommand{\eqn}{\begin{eqnarray}}
\newcommand{\eqnx}{\end{eqnarray}}
\newcommand{\ben}{\begin{eqnaray}}
\newcommand{\een}{\end{eqnarray}}
\newcommand{\f}[2]{\frac{#1}{#2}}
\newcommand{\ra}{\longrightarrow}
\newcommand{\GG}{{\cal G}}
\renewcommand{\AA}{{\cal A}}
\newcommand{\GR}{G(z)}
 \newcommand{\MM}{{\cal M}}
\newcommand{\BB}{{\cal B}}
\newcommand{\ZZ}{{\cal Z}}
\newcommand{\DD}{{\cal D}}
\newcommand{\HH}{{\cal H}}
\newcommand{\RR}{{\cal R}}
\newcommand{\GT}{{\cal G}_1 \otimes {\cal G}_2^T}
\newcommand{\GGb}{\bar{{\cal G}}^T}
\newcommand{\Du}{{\cal D}_1}
\newcommand{\Dl}{{\cal D}_2}
\newcommand{\zb}{\bar{z}}
\newcommand{\trqq}{\tr_{q\bar{q}}}
\newcommand{\arr}[4]{
\left(\begin{array}{cc}
#1&#2\\
#3&#4
\end{array}\right)
}
\newcommand{\arrd}[3]{
\left(\begin{array}{ccc}
#1&0&0\\
0&#2&0\\
0&0&#3
\end{array}\right)
}
\newcommand{\tr}{\mbox{\rm tr}\,}
\newcommand{\One}{\mbox{\bf 1}}
\newcommand{\pauli}{\sg_2}
\newcommand{\corr}[1]{\la{#1}\rangle}
\newcommand{\br}[1]{\overline{#1}}
\newcommand{\phib}{\br{\phi}}
\newcommand{\psib}{\br{\psi}}
\newcommand{\lm}{\lambda}
\newcommand{\ksi}{\xi}

\newcommand{\Gb}{\br{G}}
\newcommand{\Vb}{\br{V}}
\newcommand{\Gm}{G_{q\br{q}}}
\newcommand{\Vm}{V_{q\br{q}}}

\newcommand{\ggd}[2]{\GG_{#1}\otimes\GG^T_{#2}\Gamma}
\newcommand{\noi}{\noindent}
	
\chapter{NEW DEVELOPMENTS IN NON-HERMITIAN RANDOM MATRIX MODELS $^*$}

\author{Romuald A. Janik,\refnote{1} Maciej A. Nowak,\refnote{1,2,3} 
G\'{a}bor Papp \refnote{2,4} and Ismail Zahed\refnote{5}}

\affiliation{\affnote{1}Department of Physics, Jagellonian University,
30-059 Krak\'{o}w, Poland\\
\affnote{2}GSI, Planckstr. 1, D-64291 Darmstadt, Germany\\
\affnote{3}Institut f\"{u}r Kernphysik, TH Darmstadt, D-64289
Darmstadt,
Germany\\
\affnote{4}Institute for Theoretical Physics, E\"{o}tv\"{o}s
University,
Budapest, Hungary\\
\affnote{5}Department of Physics, SUNY, Stony Brook, NY 11794, USA}

\section{INTRODUCTION}

\footnotetext{$^*$ Talk presented by 
MAN at the NATO Workshop ``New Developments in Quantum Field Theory'',
June 14-20, 1997, Zakopane, Poland.}

Random matrix models provide an interesting framework for modeling a number of 
physical phenomena, with applications ranging from atomic physics to quantum
gravity~\refnote{\cite{WEIDNEW,ZINNJUSTIN}}. In recent years, non-hermitian
random matrix models have become increasingly important in a number of quantum
problems~\refnote{\cite{SOMMERSREV,UPDATENH}}. A variety of methods have been 
devised to calculate with random matrix models. Most prominent perhaps  are 
the Schwinger-Dyson approach \refnote{\cite{ZINNJUSTIN}} and the 
supersymmetric method \refnote{\cite{EFETOV}}. In the case of Non-hermitian 
Random Matrix Models (NHRMM) some of the standard techniques fail or are
awkward.

In this talk we go over several new developments regarding the 
techniques~\refnote{\cite{NONHER,DIAG}} for a large class of non-hermitian 
matrix models with unitary randomness (complex random numbers).
 In particular, we discuss\\ 
\noindent
(a) - A diagrammatic approach based on a $1/N$ expansion \\
(b) - A generalization of the addition theorem (R-transformation)\\
(c) - A conformal transformation on the position of pertinent singularities\\
(d) - A `phase' analysis using appropriate partition functions\\
(e) - A number of two-point functions and the issue of universality.

Throughout, we will rely on two standard examples:
a non-hermitian gaussian random matrix model (Ginibre-Girko 
ensemble~\refnote{\cite{GINIBRE}}),  and a chiral gaussian random 
matrix model in the presence of a constant non-hermitian part
~\refnote{\cite{STEPHANOV}}. The first ensemble being a text-book
example will allow for a comparison of our methods to more conventional ones, 
the second ensemble will show the versatility of our approach to new problems
with some emphasis on the physics issues. Further applications will be briefly 
mentioned.


\section{DIAGRAMMATIC EXPANSION AND SPONTANEOUS BREAKDOWN OF HOLOMORPHY}

The fundamental problem in random matrix theories is to find the
distribution of eigenvalues in the large N (size of the matrix $\MM$) limit.
According to standard arguments, 
the eigenvalue distribution is easy to reconstruct from the
discontinuities
of the Green's function
\be
G(z)=\frac{1}{N} \left< {\rm Tr}\, \frac{1}{z-\MM}\right>
\label{green}
\ee
where averaging is done over the ensemble of $N \times N$ random
matrices
generated with probability 
\be
P(\MM)=\frac{1}{Z}e^{-N {\rm Tr} V(\MM)} .
\label{probab}
\ee
To illustrate our diagrammatic 
 arguments let us first consider the well known case of a 
 random hermitian ensemble with Gaussian distribution.

\subsection{Hermitian diagrammatics}
We  use the diagrammatic notation introduced by \refnote{\cite{BZ}},
borrowing on the standard large $N$ diagrammatics for QCD
\refnote{\cite{THOOFT}}. 
Consider the 
 partition function
\be
Z=< \det (z-H) > = \int d\psi d\psi^{\dagger}  dH e^{-{\cal L}}
 e^{-\frac{N}{2} {\rm
Tr} H^2}
\label{partition}
\ee
with  a ``quark'' Lagrangian ${\cal L}$
\eqn
{\cal L} =\bar{\psi}_a ( z {\bf 1}_a^b - H^b_a )\psi^b \,, 
\label{0plus0}
\eqnx
where  $H$ is a hermitian random matrix with Gaussian weight (
the width of the Gaussian we set to 1). We will refer to 
$\psi$ as a ``quark" and to $H$ as a ``gluon". The ``Feynman 
graphs'' following from (\ref{0plus0}) allow only for the flow of  ``color'' 
(no momentum), since (\ref{0plus0}) defines a field-theory in $0+0$ 
dimensions.
The names ``quarks'',
``gluons'',
``color'' {\it etc.} are used here in a figurative sense, without any connection
to QCD.   
The ``quark'' and ``gluon'' propagators (double line notation) are shown in 
Fig.~\ref{fig.rules}. 

\begin{figure}[tbp]
\centerline{\epsfxsize=80mm \epsfbox{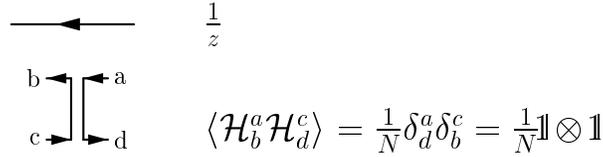}}
\caption{Large $N$ ``Feynman" rules for ``quark'' and ``gluon" propagators.}
\label{fig.rules}
\end{figure}

Introducing the irreducible self energy $\Sigma$, the Green's function reads
\eq
G(z)=\f{1}{N}{\rm Tr}\, \f{1}{z-\Sigma} = \frac{1}{z-\Sigma}\,.
\label{e2}
\eqx

In the large $N$ limit the equation for the self energy $\Sigma$
 follows from 
resumming the rainbow diagrams of Fig.~\ref{fig.rainbow}.
All other diagrams (non-planar and ``quark'' loops) 
are subleading in the large $N$ limit. 
The consistency equation (``Schwinger-Dyson'' equation 
of Fig.~\ref{fig.pastur}) reads
\eq
\Sigma=G .
\label{SD}
\eqx
Equations (\ref{e2}) and (\ref{SD}) give immediately 
$G(z-G)=1$,  so the normalizable solution for the Green's function reads 
\be
G(z)=\frac{1}{2}(z-\sqrt{z^2-4})
\label{semi1}
\ee
which, via the discontinuity (cut)  leads to Wigner's semicircle
for the distribution of the eigenvalues for hermitian random matrices
\be
\nu(\lambda)=\frac{1}{2\pi}\sqrt{4-\lambda^2}.
\ee
\begin{figure}[tbp]
\centerline{\epsfxsize=115mm \epsfbox{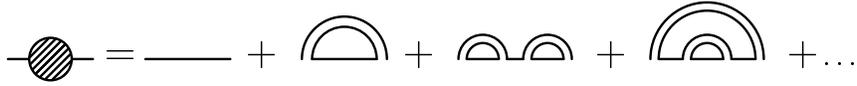}}
\caption{Diagrammatic expansion of Green's function (\protect\ref{green})
for Gaussian randomness.}
\label{fig.rainbow}
\end{figure}

\begin{figure}[tbp]
\centerline{\epsfxsize=40mm \epsfbox{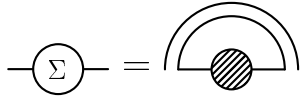}}
\caption{Schwinger-Dyson equation.}
\label{fig.pastur}
\end{figure}

\subsection{Non-hermitian diagrammatics}

If we were to use non-hermitian matrices in the resolvent (\ref{green}), then 
configuration by configuration, 
the resolvent displays poles that are scattered
around $z=0$ in the complex $z$-plane. In the large $N$ limit, the poles 
accumulate in general on finite surfaces (for unitary matrices on circles), 
over which the resolvent is no longer 
holomorphic. The (spontaneous) breaking of holomorphic symmetry follows from 
the large $N$ limit. As a result $\partial G/\partial\zb\neq 0$ on the 
nonholomorphic surface, with a finite eigenvalue distribution. In this section
we will set up the diagrammatic rules for investigating non-hermitian random 
matrix models. In addition to the ``quarks'' we introduce
``conjugate quarks'', defined by the $0+0$ dimensional Lagrangian
\eq
{\cal L}_0 = \psib(z-\MM )\psi+\phib(\zb-\MM^{\dagger})\phi .
\label{21}
\eqx
For hermitian matrices, ``quarks" $\psi$ and ``conjugate-quarks" $\phi$
decouple in the ``thermodynamical'' limit ($N\to\infty$). Their 
respective resolvents follow from (\ref{21})
and do not `talk' to each other. 
They are holomorphic (anti-holomorphic) functions modulo cuts on the
real  axis.  For non-hermitian matrices, this is not the case in the large 
$N$ limit. The spontaneous breaking of holomorphic symmetry in the large $N$ 
limit  may be probed in the $z$-plane by adding to (\ref{21})
\eq
{\cal L}_B = \lm(\psib\phi+\phib\psi)
\label{22}
\eqx
in the limit $\lm\ra 0$. The combination ${\cal L}_0+{\cal L}_B$ will be used 
below as the  non-hermitian analog of the Lagrangian   
(\ref{0plus0}).

{}From (\ref{21},\ref{22}) we define the matrix-valued resolvent through 
\be
\hat{{\cal {G}}}=\arr{{\cal G}_{qq}}{{\cal G}_{q\overline{q}}}{{\cal G}_{\overline{q}q}}
{{\cal G}_{\overline{q}\overline{q}}}
= \left\langle \arr{z-\MM}{\lambda}{\lambda}{\zb 
-\MM^{\dagger}}^{-1}\right\rangle
\label{19}
\ee
where the limit $N\rightarrow\infty$ is understood before $\lambda\rightarrow 
0$.
The ``quark'' spectral density follows 
from Gauss law~\refnote{\cite{SOMMERS88}},
\eq
\nu (z, \zb) = \frac 1{\pi} \partial_{\zb} \,\, G(z,\zb ) =
\frac 1{\pi N} \partial_{\zb} \,\, {\rm Tr}_N\,\,{\cal G}_{qq}
\label{20}
\eqx
which is the distribution of eigenvalues of $\MM$. 
For hermitian $\MM$, (\ref{20}) is valued on the real axis. As 
$\lambda\rightarrow 0$, the block-structure decouples, and we are left
with the original resolvent. For $z\rightarrow +i0$, the latter is just a 
measurement of the real eigenvalue distribution, as shown before in the
case
of Gaussian hermitian ensemble.
 For non-hermitian $\MM$,
as $\lambda\rightarrow 0$, the block
structure does not decouple, leading to a nonholomorphic resolvent for certain
two-dimensional domains on the $z$-plane. Holomorphic
separability of (\ref{21}) 
is spontaneously broken in the large $N$ limit. 
For more technical details we refer to the original work~\refnote{\cite{DIAG}}.
Similar construction has been proposed recently 
in ~\refnote{\cite{ZEENEW1}}.

\subsection{Examples}

 Consider first the Ginibre-Girko ensemble, {\it i.e.} the case of 
complex matrices with the measure
\eqn
<\ldots> =\int [d\MM] \exp \left[ -\frac{N}{(1-\tau^2)}
	{\rm tr}(\MM\MM^{\dagger}- \tau {\rm Re} \MM\MM)
\right] \,.
\label{GUEtwisted}
\eqnx
The ``gluon'' propagators read 
\eqn
\la|\MM_{ab}|^2\rangle=\f{1}{N} \qquad \,
	\la\MM_{ab}\MM_{ba}\rangle=\f{\tau}{N} 
\label{corrcomplex}
\eqnx
corresponding to hermitian ($\tau=1$), anti-hermitian ($\tau=-1$) or
general complex ($\tau=0$) matrix theory.

{}From (\ref{21},\ref{22}) we note that there are two kinds of ``quark''
propagators ($1/z$ for ``quarks'' $\psi$ and $1/\bar{z}$ for
``conjugate-quarks''$\phi$, where both can be incoming and outgoing).
The relevant ``gluonic'' amplitudes correspond now to
Fig. \ref{fig.eightglue}a--\ref{fig.eightglue}d,
where the (c,d)  contribution corresponds to twisting the lines
with a ``penalty factor'' $\tau$. 
\noindent
\begin{figure}[tbp]
\centerline{\epsfxsize=65mm \epsfbox{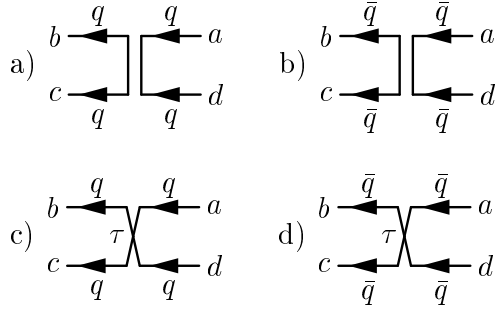}}
\caption{All ``gluonic'' amplitudes for complex gaussian matrices.}
\label{fig.eightglue}
\end{figure}

The equation for the one particle irreducible (1PI) self-energy  follows
from Figs.~\ref{fig.eightglue}--\ref{fig.twistedpastur} in the form
\eqn
\arr{\Sg_1}{\Sg_2}{\Sg_3}{\Sg_4}&=&
\f{1}{N}\tr_{N}\arr{\GG_{qq}}{\GG_{q\bar{q}}}
	{\GG_{\bar{q}q}}{\GG_{\bar{q}\bar{q}}}
\circ \arr{\tau}{1}{1}{\tau} \nonumber \\
&=&\f{1}{N}\tr_{N}
\arr{z-\Sg_1}{\lm-\Sg_2}{\lm-\Sg_3}{\zb-\Sg_4}^{-1}
\circ
\arr{\tau}{1}{1}{\tau}\,.
\label{24}
\eqnx
Here the trace is meant component-wise (block per block),
and the argument of the trace is the dressed 
propagator. The operation $\circ$ is {\em not} a matrix multiplication,
but a simple multiplication between the entries in the corresponding positions.
Here ${\rm tr}_N$ is short for the trace on the $N\times N$ block-matrices. 

Each of the entries has a diagrammatical interpretation, in analogy 
to the hermitian case. For example, the equality for 
the upper left corners  of matrices in (\ref{24})
is represented diagrammatically on Fig.~\ref{fig.twistedpastur}.
The first  graph on the r.h.s. in this figure 
does not influence the
``quark-quark'' interaction - it corresponds to the double line with a
twist,
therefore, as a non-planar one,  is subleading.
 However, this
twist could be compensated by the twisted part of the propagator 
coming from the second correlator (\ref{corrcomplex}), thereby 
explaining the factor $\tau$ in the upper left corner of (\ref{24}).
Other entries in (\ref{24}) follow from Fig.~\ref{fig.eightglue} by inspection.

\begin{figure}[tbp]
\centerline{\epsfxsize=90mm \epsfbox{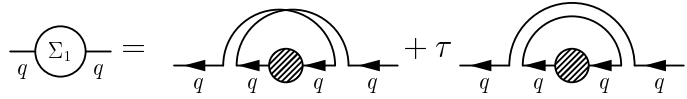}}
\caption{Self-energy equation for  non-hermitian matrices.}
\label{fig.twistedpastur}
\end{figure}

The ``quark'' one-point function is now
\eq
G(z , \zb )=\f{1}{N}{\rm tr_N }\,\, \GG_{qq}= ({\zb-\Sg_4})/{det} \,.
\label{25}
\eqx
It follows that $\Sg_2=\Sg_3$, with
 \eqn
det \cdot \Sg_1&=&{\tau(\zb-\Sg_4)} \label{e.s1}\\
det \cdot \Sg_4&=&{\tau(z-\Sg_1)} \label{e.s4}\\
det \cdot  \Sg_2&=&\Sg_2-\lm \,,
\label{26}
\eqnx
where $det=(z-\Sg_1)(\zb-\Sg_4)-(\lm-\Sg_2)^2$. Substituting
$r=\Sg_2-\lm$ in the last relation in (\ref{26}) yields the
equation
\eq
((z-\Sg_1)(\zb-\Sg_4)-r^2)(r+\lm)=r \,.
\label{27}
\eqx
For $\lm=0$, the solution with $r=0$ is holomorphic while that with $r\neq 0$
is nonholomorphic. In the holomorphic case, $\Sigma_1 (z-\Sigma_1)=\tau$, and 
the resolvent is simply
\eq
G(z)=\f{z \mp\sqrt{z^2-4\tau}}{2\tau}
\label{31}
\eqx
where the upper sign corresponds to the solution with 
the pertinent asymptotics. 
In the nonholomorphic case, $G(z, \zb) =\zb -\Sigma_4$, with
\eq
G(z , \zb)=\f{\zb-\tau z}{1-\tau^2}
\label{29}
\eqx
in agreement with \refnote{\cite{SOMMERS88}}. The boundary between the holomorphic and 
nonholomorphic solution follows from the condition $\Sigma_2=0$ imposed
for the nonholomorphic solution, here this is equivalent to 
$|G(z, \zb )|^2= 
|G(z)|^2=1$, that is
\eq
\f{x^2}{(1+\tau)^2}+\f{y^2}{(1-\tau)^2}=1
\label{34}
\eqx
which is an ellipse in the complex plane. Inside (\ref{34}) the solution is 
nonholomorphic 
and outside it is holomorphic. 
The case investigated by Ginibre \refnote{\cite{GINIBRE}} follows 
for $\tau=0$. It is pedagogical to compare our method of solving
this problem to the one coming from 
interpreting
the Ginibre ensemble as a two-matrix ($\HH \sim H_1 +iH_2$) model.

As a second example,  we consider 
the Chiral Random Matrix model, which 
 got recently some attention as a schematic
model for spontaneous breakdown of the chiral symmetry. 
Here we consider for simplicity the Gaussian version 
of such model in the presence of a non-hermitian 
part, ``chemical potential'' $\mu$,
as suggested by Stephanov~\refnote{\cite{STEPHANOV}}.
 The non-hermitian character comes from the
property
of Dirac matrices in Euclidean space.
The form of the determinant
 stems from the constant 
mode sector of the massless and chiral Dirac operator at finite chemical 
potential~\refnote{\cite{MULAT}}. 
The corresponding partition function reads
\be
Z=<\det (z-\mu \gamma -\MM)>
\ee
where
\eq
\gm=i\gamma_0 = \arr{0}{-1}{1}{0}\,\qquad{\rm and}\qquad
	\MM=\arr{0}{H}{H^\dagger}{0} \,.
\label{43}
\eqx
The only  novel features come from the ``chiral character'' of the matrix
$\MM$, 
{\it i.e.} the fact that it anticommutes with the $\gamma_5 \equiv {\rm diag}
({\bf 1}_N, -{\bf 1}_N)$. 
Due to this fact, the ``gluon'' propagator $\DD$ inherits the block structure
which in the tensor notation (see Fig.~1) reads 
\be
\DD=\frac{1}{N} (\gamma_+ \otimes \gamma_- + \gamma_- \otimes \gamma_+)
\label{chiralglue}
\ee
with $\gamma_{\pm}=({\bf 1}_N \pm \gamma_5)/2$ 
and the bare ``quark'' propagator $1/z$ gets modified 
to $(z-\mu \gamma)^{-1}$.
As a result, the 1PI self-energy equations in the planar approximation are given by 
\eq
\arr{\Sg_1}{\Sg_2}{\Sg_3}{\Sg_4}=\frac 1N \tr_{N}
	\underbrace{\arr{z-\mu\gm-\Sg_1}{\lm-\Sg_2}{\lm-\Sg_3}%
	  {\zb+\mu\gm-\Sg_4}^{-1}}_{\mbox{\Large $\hat{\GG}$}}
\circ
\arr{\DD}{\DD}{\DD}{\DD}
\label{44}
\eqx
where $\DD$ is the ``gluon'' propagator (\ref{chiralglue}), and 
$\Sg_i$ are diagonal $2N\times 2N$ matrices. Inverting in (\ref{44})
with respect to the ``quark-conjugate quark''
indices gives, after some elementary
algebra,
two kinds of  solutions:

(i)  A nonholomorphic solution ($\Sigma_2=\Sigma_3 \neq 0$)
( ``quark-quark'' resolvent)
\eq
G (z, \zb) = \f{1}{2N} \tr_{N}\ \GG_{qq} =\f{x}{2}-iy-\f{1}{2}\f{iy}{y^2-\mu^2}
\label{50}
\eqx
where $z=x+iy$, a  result first derived in \refnote{\cite{STEPHANOV}} 
using different arguments.

(ii) For $\Sigma_2=\Sigma_3=0$ we recover the holomorphic
solution~\refnote{\cite{STEPHANOV,USMUX}}, $\Sigma_1(z)=G(z)\One$,
$\Sigma_4=\Sigma_1^\dagger$, with $G(z)$ fulfilling
the cubic Pastur-like equation
\eq
G^3(z) -2zG^2(z) + (z^2 +\mu^2 +1)G(z)-z=0.
\label{cubicpasturmux}
\eqx

Note that in the case of this example 
the standard techniques of multi-matrix models do not apply.

\section{ADDITION LAWS}

The concept of addition law for hermitian ensembles 
was introduced in the seminal work by
Voiculescu\refnote{\cite{VOICULESCU}}. In brief, Voiculescu 
proposed the additive transformation (R transformation), which linearizes
the convolution of non-commutative matrices, alike the 
logarithm of the Fourier transformation  for the convolution 
of  arbitrary functions. This method is 
 an important shortcut to obtain the equations for the Green's
functions
for a sum of matrices, starting from the knowledge of the Green's functions
of individual ensembles of matrices. This formalism was reinterpreted
diagrammatically by 
Zee~\refnote{\cite{ZEE}}, who introduced the concept of Blue's function.
Let us consider the problem of finding the Green's function 
of a sum of two  independent (free~\refnote{\cite{VOICULESCU}})
 random matrices ${\cal M}_1$ and ${\cal
M}_2$,
provided we know the Green's functions of each of them 
separately.
First, we note that the 1PI self-energy $\Sigma$ can be always expressed
as a function of $G$ itself and {\em not of z} as usually done in the 
textbooks. For the Gaussian randomness, $\Sigma_H(G)=G$ (see (\ref{SD})). 
Second, we note that the graphs contributing to the self-energy
$\Sigma_{1+2}(G)$ split into two classes, belonging
to $\Sigma_1(G)$ and $\Sigma_2(G)$, due to the independence of probabilities
$P(M_1)$ and $P(M_2)$ and large $N$ (planar) limit.
Therefore
\be
 \Sigma_{1+2}(G)= \Sigma_1(G) + \Sigma_1(G) .
\label{sumgg}
\ee
Note that such a formula is not true if the energies are expressed as 
functions of $z$. Voiculescu R transformation is nothing but 
$R(u)\equiv \Sigma[G(u)]$.
The addition (\ref{sumgg}) reads, for an arbitrary complex $u$,
$R_{1+2}(u)=R_1(u)+R_2(u)$. The R operation forms and abelian group.
The Blue's function, introduced by Zee~\refnote{\cite{ZEE}}, is simply 
\be
B(G)=\Sigma(G)+G^{-1} .
\label{blue}
\ee
Therefore, using the identity $G(z)=(z-\Sigma)^{-1}$, we see that the
Blue's function is the functional inverse of the Green's function
\be
B[G(z)]=z
\label{blueinv}
\ee
and the addition law for Blue's functions reads
\be
B_{1+2}(z)=B_1(z)+B_2(z)-\frac{1}{z} .
\label{addblue}
\ee
The algorithm of addition is now surprisingly simple:
Knowing $G_1$ and $G_2$, we find (\ref{blueinv}) $B_1$ and $B_2$.
Then we find the sum $B_{1+2}$ using (\ref{addblue}), and finally, 
get the answer $G_{1+2}$, by reapplying (\ref{blueinv}). 
Note that the method treats on equal footing the Gaussian and
non-Gaussian 
ensembles, provided that the measures $P_1$ and $P_2$ are independent
(free).

The {\it naive} extension of this  algorithm fails
completely for the non-hermitian matrices. It is not {\it a priori}
puzzling
- the underlying mathematical reason for (\ref{addblue})
 is the holomorphy of the 
hermitian Green's function, not fulfilled for the case of NHRMM,
 as demonstrated in the previous section. 
However, since we managed to extend the  diagrammatical analysis to the NHRMM, 
it is still possible to generalize
 the addition formula using the parallel
diagrammatic reasoning like in the hermitian case.  
The generalization amounts to consider the matrix-valued Green's function
(\ref{19}). 
The generalized Blue's function~\refnote{\cite{NONHER,DIAG}}
 is now a matrix valued function of
a $2 \times 2$ matrix variable defined by
\be
\BB(\GG)=\ZZ=\left( \begin{array}{cc} z & \lambda \\ \lambda & \bar{z}
                    \end{array} \right) . 
\ee
where $\lambda$ will be eventually set to zero. This is equivalent
to the definition in terms of the self-energy matrix
\be
\BB (\GG )=\Sigma+\GG^{-1}
\label{genblue}
\ee
where $\Sigma$ is a $2 \times 2$ self energy matrix expressed as a
function of a matrix valued Green's function. The same diagrammatic
reasoning
as before leads to the addition formula for the self-energies and
consequently
for the addition law for generalized Blue's functions
\be
\ZZ  =\BB_1(\GG)+\BB_2(\GG) -\GG^{-1} .
\label{genadd}
\ee
The power of the addition law for NHRMM stems from the fact that it
treats
Gaussian and non-Gaussian randomness on the same footing~\refnote{\cite{ZEENEW2}}.

\subsection{Example}
Let us consider for simplicity  complex random Gaussian matrices
($\tau=0$), which we rewrite as the sum $H_1+ i H_2$, with $H_1,
H_2$ hermitian. 
The generalized Blue's function  for hermitian $H_1$ follows explicitly from
\be
\Sigma^{(1)} \equiv 
\left( \begin{array}{cc}  \Sigma^{(1)}_1 & 
\Sigma^{(1)}_2 \\ \Sigma^{(1)}_3 & \Sigma^{(1)}_4   \end{array} \right)
= 
 \hat{\GG} \circ 
 \left( \begin{array}{cc} 1 & 1 \\1& 1 \end{array} \right) =\hat{\GG}
\label{one}
\ee
a matrix analog of (\ref{SD}).
The generalized Blue's function for anti-hermitian $iH_2$ follows from
\be
\Sigma^{(2)}=\hat{\GG} \circ \left( \begin{array}{rr} -1 & 1 \\1& -1 \end{array} \right) 
\label{two}
\ee
where the entries reflect the antihermicity (set $\tau=-1$ in (\ref{24})).
It is a straightforward exercise to check  that the matrix equation
(\ref{genadd}) with the  generalized Blue's functions $\BB_1$ and $\BB_2$, 
corresponding to (\ref{one}) and (\ref{two}),
reproduces   two types of solutions (\ref{31}) and (\ref{29}) 
as well as the 
equation for the boundary ( here the circle) separating them on 
the $z$ plane.

\section{CONFORMAL MAPPINGS}

The   
existence of the nonholomorphic and as well holomorphic 
domains in the case of two solutions of NHRMM provides a powerful
way
to evaluate the supports for the level densities of NHRMM. 
The envelopes of these supports (supports form two-dimensional islands)
 can be derived
very generally using a conformal transformation that maps the cuts
of the hermitian ensemble onto the boundaries of its non-hermitian
analog.

Let us consider the case  where a Gaussian random and hermitian matrix
$H$ is added to an arbitrary matrix $M$. 
The addition law says
\be
R_{H+M}(u)=R_H(u)+R_M(u)=u+R_M(u).
\label{hm}
\ee
where we have used  explicitly that for Gaussian $R_H(u)=u$.
 Now, if we were to note that in the {\em holomorphic} domain
the R transformation for the Gaussian  anti-hermitian ensemble
is $R_{iH}(u)=-u$, we read\footnote{Note that anti-hermitian Gaussian
nullifies in the holomorphic domain 
the hermitian Gaussian in the sense of the group property of 
additive transformation $R$, i.e. $R_{H}+R_{iH}={\bf 0}$.} 
\be
R_{iH+M}(u)=R_{iH}(u)+R_M(u)=-u+R_M(u).
\label{ihm}
\ee
These two equations yield
\be
B_{iH+M}(u)=B_{H+M}(u)-2u
\label{sumBLet}
\ee
where we have used the  relation $B(u)=R(u)+1/u$. 
Substituting $u\rightarrow G_{H+M}(z)$ we can rewrite (\ref{sumBLet}) as
\be
B_{iH+M}[G_{H+M}(z)]=z-2G_{H+M}(z) \,.
\label{map1}
\ee
Let $w$ be a point in the complex plane for which
$G_{iH+M}(w)=G_{H+M}(z)$.
Then 
\be 
w=z-2G_{H+M}(z)\,.
\label{map2}
\ee
Equation (\ref{map2}) provides a conformal transformation mapping
the {\it holomorphic} domain of the ensemble $H+M$ ({\it i.e.}
the complex plane $z$ minus cuts) onto the {\it holomorphic} domains of 
the ensemble
$iH+M$, {\it i.e.} the complex plane $w$ minus the ``islands'',  defining 
in this way the 
support of the eigenvalues.
\subsection{Examples}
Consider the case of ``summing'' two Hermitian random Gaussian ensembles,
{\it i.e.} consider the Hamiltonian $H=H_H+gH_H$, where $g$ is some arbitrary
coupling. The sum constitutes of course the Gaussian ensemble, 
and the spectrum follows
from the properties of the $R$ function $R_{H+gH}(z)=(1+g^2)z$, or
equivalently,
Green's function
\be
G(z)=\frac{1}{2(1+g^2)}[z-\sqrt{z^2-4(1+g^2)}]
\label{gaussadded}
\ee
 {\it i.e.} the support of the eigenvalues forms the interval (cut)
${\cal{I}}=[-2\sqrt{1+g^2},+2\sqrt{1+g^2}]$. 
According to (\ref{map2}), we can map the interval ${\cal{I}}$ onto the
boundary
delimiting the holomorphic domain of the {\it non-hermitian} ensemble $\HH=
H_H+i\gamma H_H$,  ($g\rightarrow i\gamma$), that is 
\be
w=\frac{1}{1+g^2}[g^2z+\sqrt{z^2-4(1+g^2)}]
\label{island}
\ee
with $z=t\pm i0$ and $t$ in ${\cal{I}}$. Equation (\ref{island}) spans an 
ellipsis with axes $2/\sqrt{1+g^2}$ and $2g^2/\sqrt{1+g^2}$.
For $g^2=1$ the ellipsis is just the Ginibre's circle.

A similar construction and an identical mapping (\ref{map2}) gives the
support of 
the eigenvalues in the case of a schematic chiral Dirac operator with
chemical potential. Let us first consider the case
when $\mu=i\epsilon$, {\em i.e.} the case when the ensemble is hermitian.
In this case, the resulting Green's function is known to fulfill the 
so-called Pastur equation (random gaussian plus  deterministic
hermitian Hamiltonian 
$E$,  here with $N/2$ levels $\epsilon$ and $N/2$ levels 
 $-\epsilon$)
\be
  G(z)=\frac{1}{z-G(z)-\epsilon}+\frac{1}{z-G(z)+\epsilon}
\label{Pastur}
\ee
encountered in many areas of physics~\refnote{\cite{PASTUR}}. 
This is exactly equation (\ref{cubicpasturmux}) 
with the formal replacement $\mu^2 \rightarrow
-\epsilon^2$. At a particular value of the deterministic parameter
$\epsilon=1$, 
the single cut supporting the spectrum of the hermitian ensemble
splits into two-arc  support, manifesting therefore a structural change 
in the spectral properties, hence a ``phase transition''. 
The spectral properties
of the non-hermitian model, with chemical potential $\mu$, follow from
the mapping (\ref{map2}), but with $G_{H+M}$ replaced by an  appropriate
branch
of the cubic Pastur equation~(\ref{Pastur}). In particular, at the value 
$\mu^2=1$ the spectrum demonstrates the structural change - an island
splits into two disconnected mirror islands (see Fig.\ref{fig.maps}).

\begin{figure}[tbp]
\centerline{\epsfxsize=55mm \epsfbox{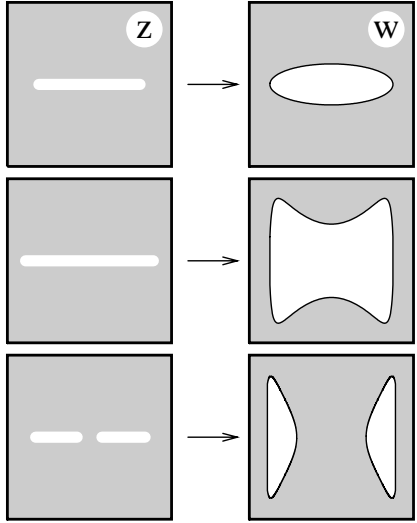}}
\caption{Conformal mappings for the case of Ginibre-Girko ensemble with 
$\tau=0.5$ (upper),
non-hermitian chiral ensemble with chemical potential $\mu^2=0.8$
(middle) 
and $\mu^2=1.2$ (lower). The shaded regions represent the holomorphic
domains.}
\label{fig.maps}
\end{figure}

\section{TWO-POINT FUNCTIONS}

To probe the character of the correlations between the eigenvalues of 
non-hermitian random matrices, either on their holomorphic or nonholomorphic 
supports, it is relevant to investigate two-point functions.
For example, a measure of
the breaking of holomorphic symmetry in the eigenvalue distribution 
is given by the connected two-point function or correlator
\eq
N^2G_c(z,\bar{z})=\corr{\left| \tr \f{1}{z-\MM}\right|^2}_c
\label{63}
\eqx
where the $z$ and $\zb$ content of the averaging is probed simultaneously.
The correlation function (\ref{63}) will be shown to diverge
precisely on the nonholomorphic support of the eigenvalue distribution,
indicating an accumulation in the eigenvalue density.
In the conventional language of ``quarks'' and ``gluons'', 
(\ref{63}) is just the correlation function between ``quarks" and their
``conjugates". A divergence in (\ref{63}) in the $z$-plane reflects 
large fluctuations between the eigenvalues of the non-hermitian operators
on finite $z$-supports, hence their ``condensation". 

It was shown in \refnote{\cite{AMBJORN}} and \refnote{\cite{BREZIN}}
 that
 for hermitian
matrices (with $\zb\to w$) the fluctuations in connected and smoothened
two-point functions satisfy the general lore of macroscopic
universality. 
This means that all smoothened correlation functions are universal and 
could be classified by the support of the spectral densities, 
independently of the specifics of the random ensemble and genera in the 
topological expansion (see \refnote{\cite{AKEMANN}} for a recent discussion). 

In the case on NHRMM the generalized two-point correlator reads~\refnote{\cite{DIAG}}
\eq
\label{l.form}
N^2G_c(z,w)=-\partial_z\partial_w \log \det(1-\GT\Gamma) \,.
\eqx
Here the logarithm is understood as a power series expansion.
Equation (\ref{l.form}) is valid for Gaussian ensembles and, 
in the general case, up to factorizable corrections in the sense
of \refnote{\cite{BZ}}. 
The operator $\GT\Gamma$ is a tensor product of $2\times 2$ matrices
(see Fig.~\ref{fig.sector}). 
The kernel $\Gamma$ includes the details of the ``gluonic''
interactions,
depending on the particular measure. 
The tensor structure reflects  the nonholomorphic solutions.
The choice of ``isospin" in the {\it e.g.} lower fermion
line is done by choosing the appropriate derivative $\partial_w$ for the 
``quark" and $\partial_{\br{w}}$ for the ``conjugate-quark".
\begin{figure}[tbp]
\centerline{\epsfxsize=85mm \epsfbox{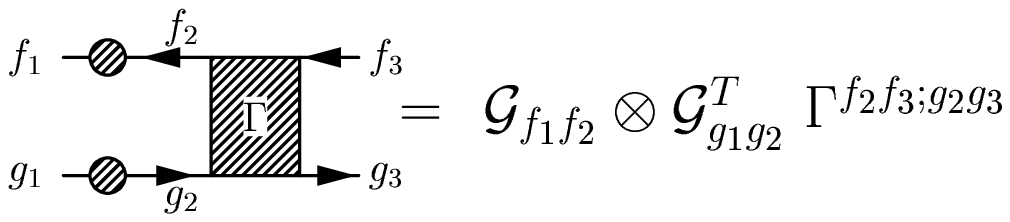}}
\caption{Two-point kernel with $f,g = q, \bar{q}$.}
\label{fig.sector}
\end{figure}

For the holomorphic solutions of NHRMM, the structure is simpler, since 
the Green's functions are  holomorphic in this case.

\subsection{Examples}
In the case of the Ginibre Girko ensemble, 
the two-point correlator in the {\it holomorphic domain} reads simply
\be
C(z,\bar{z})= -\partial_z \partial_{\bar{z}} \ln [1-G(z)G(\bar{z})] \,.
\label{girkoholo}
\ee
Indeed, in the holomorphic domain instead of the matrix valued $\GG$ we use
$G(z)$ given by (\ref{31}), and the kernel $\Gamma$ reduces to the
``quark-conjugate quark'' coupling equal to 1 (upper-right corner of the
last matrix in (\ref{24})). 
Note that the correlator (\ref{girkoholo}) diverges on the line
\be
1-|G(z)|^2=0
\label{elipsa}
\ee
therefore the ellipse (\ref{34}), confirming our statement.

For the nonholomorphic domain, the calculation is a bit more involved,
due to the explicit matrix structure of $\GG$ and $\Gamma$.
The explicit form of the matrix-valued resolvent is
\eqn
\GG = \frac{1}{1-\tau^2}
\left( \begin{array}{cc} \zb-\tau z &g_z\\g_z&z-\tau \zb \end{array}\right) 
\label{mmm}
\eqnx
with $g_z^2=|z|^2(1-\tau)^2-\tau(z+\zb)^2-(1-\tau^2)^2$.
One recognizes (\ref{29}) as the upper left corner of (\ref{mmm}).
The explicit form of the kernel is 
\be
\label{e.gamma}
\Gamma ={\rm diag} ( \tau, 1,1,\tau)
\ee
which corresponds to all possible contributions from four  graphs on
Fig.~\ref{fig.eightglue}.
After some algebra, the  determinant of 4 by 4 matrix $(1-\GT\Gamma)$ 
turns out to be
equal to $|z-w|^2$, 
giving the correlator
\eqn
N^2 G_{qq}(z,w)&=&-\f{1}{(w-z)^2} \label{93} \,.
\eqnx
In the non-hermitian chiral case the correlator in the {\it outside}
(holomorphic) region 
is calculated using the same arguments as above. 
The only minor technical complication stems from the 
chiral (block) nature of the Green's functions.
We skip the details published elsewhere~\refnote{\cite{DIAG,USMUX}},
 showing only the final result.
The determinant of $1-\GT\Gamma$ gives
\eq
\frac{(D-\mu^2)^2-|z-G|^4}{D^2}
\label{detmodul}
\eqx
where the  holomorphic G is the appropriate branch of
(\ref{cubicpasturmux})  and $D=|(z-G)/G|^2$.

The zero of the determinant in (\ref{l.form})
occurs for $(D-\mu^2)=|z-G|^2$, that is
\eq
|z-G|^2(1-|G|^2)-\mu^2 |G|^2=0
\label{e.ch1z}
\eqx
as quoted in \refnote{\cite{USMUX}}. This is exactly the equation of the
boundary 
separating the holomorphic and nonholomorphic solutions, obtained in
examples
before 
either as $\Sigma_2=0$ or as a result of conformal mapping.
 
In the hermitian case $\mu=0$ (and $\zb=w$), the determinant
in (\ref{detmodul}) is simply $(1-G^2(z) G^2(w))$ (chiral) as opposed to
$(1-G(z)G(w))$ (non-chiral). As a result, for $w=z$ and $\mu=0$,
 (\ref{detmodul}) is
\be
N^2 G(z,z)=\frac{1}{z^2 (z^2-4)^2}
\label{chiralduo}
\ee
which coincides with (5.5) in \refnote{\cite{MAKEENKO}}. 

 Note that the form of (\ref{chiralduo}) signals two kinds of {\em 
microscopic universalities}. The $1/N$ expansion 
breaks down at $z=\pm 2$ (endpoints of spectra) and $z=0$ (
``Goldstone'' pole due to the ``chiral'' nature of the ensemble).

The divergence at $z=\pm 2$ points at the 
edge universal behavior of the spectral function 
(Airy oscillations)~\refnote{\cite{AIRY}},
the divergence at $z=0$ signals the chiral microscopic 
universality~\refnote{\cite{BESSEL}}.
Unfolding the spectra at these points allows to get the explicit
universal kernels characterizing the fore mentioned universalities.

In the light of the above remarks, it is tempting to
speculate~\refnote{\cite{DIAG}}, 
that
the divergences of generalized correlators may onset some {\it new types}
of microscopic universalities present in the NHRMM. 

Note also that the relations (\ref{elipsa}),  (\ref{detmodul}) 
demonstrating the {\em functional} 
dependence of the two-point correlator on one point holomorphic 
Green's function allow for some extension
of  the macroscopic universality for NHRMM as well. 
The eventual geometric interpretation of this extension remains  an open
problem.   

Before closing this example let us present for completeness the result
of the chiral correlator in the nonholomorphic domain. 
The calculation is tedious, due to 
the fact that in the 
nonholomorphic region ``quarks" may turn to ``conjugate-quarks" and vice-versa, with 
all ``quark" species interacting with themselves, and species
appear 
in  chiral copies.
Nevertheless, the final result for the determinant
is remarkably simple:
\be
\hspace*{-8mm}
\det(1\!-\!\GG(z)\otimes\GG^T(w)\Gamma) = 
	|z\!-\!w|^2|z\!+\!w|^2
	\f{(\mu^2\!-\!(\mu^2\!-\!y^2)(\mu^2\!-\!v^2))^2\!-\!v^2y^2}{\mu^4}
	\, 
\ee
(where $y= {\rm Im}\, z, v={\rm Im}\, w$), 
suggesting perhaps the possibility of further technical developments
in the case of NHRMM.

\section{PARTITION FUNCTIONS}

\newcommand{\sg}{\sigma}
\newcommand{\al}{\alpha}
\newcommand{\PP}{{\cal P}}
\newcommand{\dz}{\partial_z}
\newcommand{\dzb}{\partial_{\zb}}
\renewcommand{\gg}{\GG\otimes \GG^T \cdot \Gamma}
\newcommand{\ggb}{\GG\otimes \br{\GG}^T \cdot \Gamma}
\newcommand{\gbgb}{\br{\GG}\otimes \br{\GG}^T \cdot \Gamma}
\newcommand{\ggg}{\det(1-\GG\otimes \br{\GG}^T \cdot \Gamma)}

We show now that the information carried 
by the one- and two-point functions is sufficient to
specify the ``thermodynamical"  potential to order ${\cal O}(1/N)$ in the
entire $z$-plane modulo isolated singularities, as we now discuss. 
Similar ideas were used in different context in
\refnote{\cite{STEELE,ITOI,AMBJORNETAL}}.

Let $Z_N$ be the partition function in the presence of an external parameter
$z$. In the $1/N$ approximation, the diagrammatic contributions to the
 partition function $Z_N$ read
\be
\log\, Z_N = NE_0 + E_1 + {\cal O} (\frac 1N )
\label{YL1}
\ee
where $E_0$ is the contribution of the ``quark" or ``conjugate quark" loop
in the planar approximation, and $E_1$ is contribution of the 
``quark-quark" loop, and so on, in the same approximation. We
 will restrict our attention to non-hermitian matrices with unitary 
randomness, in which case the non-planar corrections to $E_0$ are
of order $1/N^2$.
Hence, $E_0$ is determined by the one-point function and $E_1$ by the 
two-point functions. 

For $z$ such that  (\ref{YL1}) is real, continuous and 
nondecreasing function of the extensive parameters~\refnote{\cite{HUANG}}, 
$\log\, Z_N/N$ may be identified with the ``pressure" of the random matrix 
model. As a
result, the isolated singularities in the ``pressure" are just the ``phase"
boundaries provided that the expansion is uniform. Below we give 
examples where the ``phase" boundary is either mean-field-driven or 
fluctuation-driven. We have to distinguish two cases:
holomorphic partition functions (``unquenched'') and 
nonholomorphic partition functions, where the complex phases of the 
determinants are neglected.

\subsection{Holomorphic $Z$}

We consider the partition function
\eq
Z_N =\corr{\det (z-\MM) }=\left<\int d\psi d\psib 
e^{-\psib(z-\MM)\psi}\right> \,.
\label{X106}
\eqx
In contrast to the one- and two-point correlators discussed above, the 
determinant in (\ref{X106}) is not singular in the $z$-plane configuration 
by configuration. Hence, (\ref{X106}) is {\it a priori} holomorphic in 
$z$ (minus isolated singularities).

The one- and two-point functions on their holomorphic support may be
obtained from  
$\mbox{log}\,Z_N$ by differentiation with respect to $z$. Therefore, from 
(\ref{YL1})
\eq
E_0=\int^z\! dz'\ G(z') + const
\label{eq.E0}
\eqx
or equivalently
\be
E_0 =zG - \int\! dG\ z(G) + const
\label{trick}
\ee
after integrating by part. Note that $z(G) =B(G)$ is just the Blue's 
function~\refnote{\cite{ZEE}} of $G$.
 The constant in $E_0$ is fixed by the 
asymptotic behavior of (\ref{X106}), that is $Z_N\sim z^N$. 
The planar contribution to $E_1$ in (\ref{YL1}) follows from the
``quark-quark" wheel (two-point correlator). The final
 result for $Z_N$ is
\be
Z_N =e^{N E_0}\cdot \left(\left\{ \det(1-\gg)\right\}^{-\frac 12} + 
{\cal O} \left( \frac 1N \right) \right) \,.
\label{X107}
\ee
Note that due to the power $-1/2$, the fluctuations have ``bosonic"
character, and are dwarfed  by the ``quark" contribution
as ($1:N$) \refnote{\cite{USNJL}}.
Both $E_0$ and $E_1$ are simple functions of the resolvent on a specific
branch, as expected from generalized macroscopic universality.

We note that the partition function 
$Z_N$ through (\ref{X107}) exhibits an essential singularity in $1/N$ as 
expected from thermodynamical arguments. Assuming that the expansion 
for $\log Z_N /N$ is uniform, then $\log Z_N/N$ follows from (\ref{X107}) using
the holomorphic resolvent $G(z)$ for large $z$. The small $z$ region
follows by 
analytical continuation. However, since $G(z)$ is multi-valued (already the
simple case of the Ginibre-Girko ensemble yields two branches for the
resolvent in~(\ref{31})), the analytical
continuation is ambiguous. The ambiguity may be removed by identifying
$\log |Z_N|/N$ with some {\it generalized}
``pressure" and taking $G(z)$ so that $\log |Z_N|/N$ is maximum. 
As a result, $V_N=\log \, Z_N/N$ is piece-wise 
analytic in leading order in $1/N$ with ``cusps" at
\eq
F^{(ij)}(x,y) \equiv V_N^{(i)}(x,y) - V_N^{(j)}(x,y) = 0\,,
\label{CUSPS}
\eqx
following the transition from branch $i$ to branch $j$ of $G$.

The character of the transition in the $1/N$ approximation
can be highlighted by noting that
for any finite $N$, the partition function (\ref{X106}) 
is a complex polynomial in $z$ of degree $N$ with random coefficients. 
In large $N$,
\eq
V_N=\frac 1N \log\, |Z_N| = \frac 12 \int \, dv\,d\overline{v}\,\varrho (v, 
\overline{v})\, \log |z-v|^2 \,.
\label{INTEGRAL}
\eqx
To leading order, the distribution of 
singularities along the ``cusps" (\ref{CUSPS}) is 
\be
\varrho (z, \zb ) = \frac 1{2\pi} | \dz F |^2 \,\, \delta ( F (z, \zb) )
\label{densdel}
\ee
which is normalized to 1 in the $z$-plane. Redefining the density of 
singularities by unit length along the curve $F(z,\zb)=0$, we may rewrite
(~\ref{densdel}) as
\be
\left. \varrho(z,\zb)\right|_{F=0} = \frac 1{2\pi} |\dz F | \equiv
	\frac 1{4\pi} |G^{(i)}-G^{(j)}| \,.
\label{densun}
\ee
For $\varrho \neq 0$, the integrand
in (\ref{INTEGRAL}) is singular at $z=v$ which results into different forms
for $V_N$, hence a cusp. For $\varrho =0$, that is $\dz F =0$,
$V_N$ is differentiable. For physical 
$V_N$ (real and monotonically increasing), the points $\varrho =0$ are 
multi-critical points. At these points all $n$-point ($n\geq 2$)
functions diverge.
This observation also holds for Ising models with complex external
parameters \refnote{\cite{ROBERT}}. Assuming
macroscopic universality~\refnote{\cite{AMBJORN}} for all $n$-points ($n\geq 2$), 
we conclude that $\partial_z F =0$ means a branching point for the
resolvents, hence $\dz G =\infty$ 
or $B' (G) =0$ \refnote{\cite{ZEE}}. 
 For hermitian matrices, these conditions coincide with 
the end-points of the eigenvalue distributions \refnote{\cite{ZEE,USBLUE}}.

\renewcommand{\dz}{\partial_z}
\renewcommand{\dzb}{\partial_{\zb}}
\newcommand{\ddx}{\partial_x}
\newcommand{\ddy}{\partial_y}

\subsection{Examples}

To illustrate the above concepts, consider again first 
the Ginibre-Girko ensemble.
The resolvent in the holomorphic region satisfies (\ref{31}), so 
\be
z = \tau G+\frac{1}{G}\, .
\label{defgir}
\ee
The integration~(\ref{trick}) in $E_0$ is straightforward, and after
fixing the asymptotic behavior we obtain 
\be
Z_N  = G^{-N} e^{\frac{\tau}{2}N G^2} \left( 
(1-G^2(z)\tau)^{-\frac{1}{2}}
 + {\cal O} \left(\frac 1N\right) \right)\,.
\label{girhol}
\ee
Here $G$ is the solution of (\ref{defgir}). The pre-exponent in (\ref{girhol})
follows from (\ref{X107}) with the matrix $\GG$ replaced 
by $G$ and $\Gamma=\tau$, as seen in the 
``quark-quark" component of the vertex matrix in (\ref{24}).
Using (\ref{defgir}) we observe that the pre-exponent diverges at two 
points in the $z$-plane, $z^2 =4\tau$. At these points there is a ``phase" 
change as we now show. 

Given (\ref{girhol}), the generalized ``pressure" in leading order is
\be
V_\pm = -\frac{1}{2}\mbox{log\,}(G_\pm\overline{G}_\pm) + \frac{\tau}{4}
	(G_\pm^2+\overline{G}_\pm^2) +{\cal O} \left( \frac 1N\right)\,.
\ee
$V_\pm$ define two intersecting surfaces valued in the $z$-plane, 
for two branches $G_{\pm}$ of the solutions to (\ref{defgir}). 
The parametric equation for the intersecting curve is  
\be
F (z , \zb ) = V_+ -V_- =0 .
\label{CUTX}
\ee
As indicated above, $V_N$ is piece-wise differentiable.
Note that $F=0$ on the cut along the real axis, 
$-2\sqrt{\tau}<z<+2\sqrt{\tau}$, and from~(\ref{densun}) the density of
singularities per unit length is
\eq
\left. \varrho (z,\zb)\right|_{F=0} = \frac 1\pi 
	\frac{\sqrt{4\tau-z^2}}{2\tau} \,. 
\label{charge1}
\eqx
Along $F$, the density of singularities is semi-circle. The density
(\ref{charge1}) vanishes at the end-points $z=\pm 2\sqrt{\tau}$. This is
easily seen to be the same as $\dz G = \infty$, or $dB(G)/dG =0$ with 
$B (G) =\tau G + 1/ G$.  As noted above the term in bracket in
Eq.~(\ref{girhol}) vanishes at these points, with a diverging
``quark-quark" contribution. The transition is fluctuation-driven. 
These points may again signal the onset of scaling regions with possible universal
microscopic behavior for non-hermitian random matrix models. 
This issue
will be pursued elsewhere. At these points the $1/N$ expansion we have used 
breaks down. 

Let us move now to the chiral non-hermitian ensemble.
Elementary integration for this case leads to 
\eq
Z_N (z, \mu) =e^{N E_0}\cdot \left(\left\{ D^{-2} [(D+\mu^2)^2 -(z-G)^4]
\right\}^{-\frac 12} + 
{\cal O} \left( \frac 1N \right) \right)\,,
\label{X107ch}
\eqx
but now  $D=(z-G)^2/G^2$, and
\eq
E_0 (z, \zeta ) = G^2 +\log \f{z-G}{G}
\eqx
with the appropriate branch of holomorphic $G$ solution to 
(\ref{cubicpasturmux}).

Note that for $z=0$ and $G^2=-1-\mu^2$, the pre-exponent in (\ref{X107ch})
diverges. It also diverges at $z=z_*$ which are the zeros of
(\ref{densun}) for the present case (two zeros for small $\mu$ and
four zeros for large $\mu$), see Fig.~\ref{fig.crit}.
 Again, at these points, the $1/N$ expansion 
breaks down marking  the onset of scaling regions and the possibility of 
microscopic universality. The $z=0$ divergence is just the notorious 
``Goldstone" mode in chiral models, illustrating the noncommutativity of 
$N\rightarrow \infty$ and $z\rightarrow 0$. The rest of the arguments 
follow easily from the preceding example, in agreement with the 
``thermodynamics" discussed in~\refnote{\cite{USNJL}}.
The analytical  results 
for the nature and location of singularities
of this example were confirmed by an extensive numerical analysis of
Yang-Lee
zeroes (up to
500 digits accuracy) in~\refnote{\cite{HALASZ}}.

\subsection{Nonholomorphic $Z$}

The above analysis for the holomorphic thermodynamical potential may also be 
extended to nonholomorphic partition functions of the type 
\eq
Z_N [z, \zb ] =\corr{\det |z-\MM|^2}=\left<\int d\psi d\psib d\phi
	d\overline{\phi} 
e^{-\psib(z-\MM)\psi -\phib(\zb-\MM^{\dagger})\phi}\right> \,.
\label{X100}
\eqx
Note the important ``quenching'' of the phase of the determinant
in comparison to the holomorphic case~(\ref{X106}).
As a consequence, the two-point correlators diverge rather on the
one-dimensional boundary separating the phases (when approaching
the boundary from the holomorphic domain) then at discrete points. 

Similar reasoning as before leads to the explicit expression
\eq
Z_N [z, \zb ]\!\!=\!\!e^{N E_0}\cdot\left(\!\left\{ 
	\det(1\!\!-\!\ggb)\, |\!\det(1\!\!-\!\gg)| \right\}^{-1}
	\!\!\!+\!{\cal O}\!\left(\!\frac 1N\!\right)\!\right)
\label{X101}
\eqx
where $E_0$ comes from the solutions of 
\be
G(z, \zb)= \frac{\partial E_0(z,\zb)}{\partial z} \qquad
\overline{G}(z, \zb)= \frac{\partial E_0(z,\zb)}{\partial \zb} \,.
\ee
Note that again the contributions from the two-wheel diagrams
are of the form $1/\sqrt{{\rm det}}$ and hence ``bosonic" in character.
The result could be easily guessed without performing the calculations:
 there are two 
contributions from the ``quark-conjugate-quark" wheels (correlators) 
(square of the $1/\sqrt{\det}$  in first term in the curly bracket)
one contribution from the ``quark-quark"  wheel
 and one contribution from the ``conjugate-quark-conjugate-quark" wheel
(represented together as a second (modulus) term in the curly bracket).

Again, the partition function $Z_N$ has an essential singularity in
 $1/N$, but $\log Z_N/N$ does not. For any finite $N$, the latter 
diverges for 
\eq
\det (1-\ggb) =0
\label{LINESING}
\eqx
which defines the {\em line} of singularities, and for 
\eq
\det (1-\gg) =0
\label{DISPOINTS}
\eqx
defining  the {\em set of discrete
points}, encountered in the case of the holomorphic partition function.

\subsection{Examples}

For the Ginibre-Girko example, the line of singularities
 (\ref{LINESING})
reads 
\eq
1-|G|^2 =0 \,.
\label{ellipse}
\eqx
 The line of singularities
(\ref{ellipse}) reproduces in this case the ellipse
(\ref{34}). 
The ellipse includes the points of the ``phase"
change (see (\ref{girhol})), 
\be
1-\tau G^2(z)=0
\label{points}
\ee
i.e. the focal points $z^2=4\tau$, corresponding to (\ref{DISPOINTS}),  
connected by the interval
(\ref{CUTX}), {\it i.e.} $F=0$. 

In the case of chiral non-hermitian random model 
the condition (\ref{LINESING}) reads~\refnote{\cite{USMUX}}
\eq
 D^{-2} [(D-\mu^2)^2 -|z-G|^4] =0
\label{X107cch}
\eqx
with $D=|(z-G)/G|^2$, therefore exactly the condition (\ref{e.ch1z}). 
This line represents the boundary between the holomorphic and
nonholomorphic
solutions.
The set of discrete multi-critical points, corresponding to
(\ref{DISPOINTS})
is given by the  condition
 \eq
 D^{-2} [(D+\mu^2)^2 -(z-G)^4] =0
\label{X1077cch}
\eqx
but with  $D=[(z-G)/G]^2$, in agreement with 
(\ref{X107ch}). 
Note the crucial appearance of the modulus and the flip in the sign of
$\mu$ when comparing last two formulae. The explicit solution 
of (\ref{X1077cch}) consists on set of two or four points, (depending on
the value
of the $\mu$), being the analogs of Airy type end-points singularities
{\it and}  a single multi-critical point $z=0$, reflecting the 
chiral nature of the ensemble. 

Figure~\ref{fig.crit} shows the critical lines  and critical points
corresponding to the conditions (\ref{LINESING},\ref{DISPOINTS}) for 
Ginibre-Girko and non-hermitian chiral ensembles. End-points
singularities
 are denoted
by
``NATO stars'', chiral singularity -- by ``Zakopane sun''.

\begin{figure}[tbp]
\centerline{\epsfysize=30mm \epsfbox{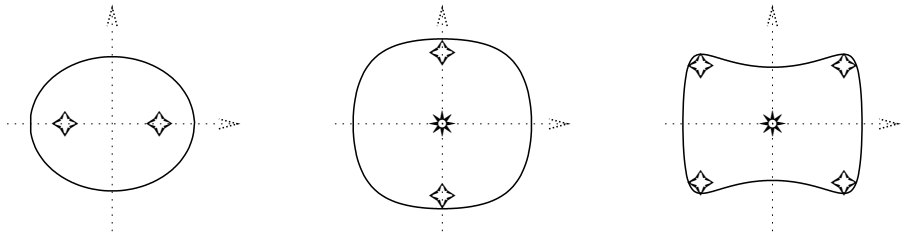}}
\caption{Critical lines (\protect\ref{LINESING}) and critical points
(\protect\ref{DISPOINTS}) for Ginibre-Girko ensemble with $\tau=0.1$  (left)
and non-hermitian chiral ensemble with small ($\mu^2=0.1$, middle) and 
large ($\mu^2=0.6$, right) 
value of chemical potential.  The solid lines and the symbols represent
the manifold on which the  $1/N$ expansion breaks down and 
signal the location of possible microscopic 
universal behavior for NHRMM.}
\label{fig.crit}
\end{figure}

The fact that the critical line in  Fig.~\ref{fig.crit}~b,c
 surrounds the multi-critical points of the unquenched partition
function,
explains the failure of quenched lattice calculations with finite 
baryonic potential. The nature of chiral restoration is masked 
by unphysical fluctuations caused by neglecting the phase of the
 determinant.
The  critical line (\ref{X107cch}) 
  exactly reproduces the shoreline of islands of
unphysical ``mixed-condensate'', obtained using the replica 
methods~\refnote{\cite{STEPHANOV}}.

\section{CONCLUSIONS}

Most of the details of 
results presented here are included in already published
papers~\refnote{\cite{NONHER,DIAG,USMUX}}.
 In this mini-review on some novel aspects of 
NHRMM, we tried to enhance the role 
of  {\em a~priori} not obvious 
connections between the hermitian and non-hermitian ensembles of
 random matrices.   
In particular, we presented several ways for providing the supports
and the density of eigenvalues for non-hermitian ensembles and the way
for calculating smoothened (wide) correlations, 
via either analogies (diagrammatic expansion, generalized Blue's
functions), 
or formal relations between the hermitian and non-hermitian ensembles
(conformal mapping, quenching/unquenching of partition function).
We used the same set of known examples to demonstrate clearly the
cross-references
between the methods, as well to exhibit the shortcomings
of the standard  treatment. 

We also speculated on some new features, like 
generalization of macroscopic universality and the possibility 
of several types of new microscopic universalities related to the critical
behavior
of various correlators in the NHRMM. 

This last issue is of great interest in light of  recent 
exciting results
in ``weakly'' non-hermitian random matrix models~\refnote{\cite{WEAKLY}}
and related  applications~\refnote{\cite{HATANONELSON}}.

\subsection{Acknowledgments}

This work was partially supported by the US DOE grant DE-FG-88ER40388,
by the Polish Government Project (KBN) grant 2P03004412
and by Hungarian grants OTKA T022931 and FKFP0126/1997.
MAN thanks the organizers of the Workshop for an opportunity to 
present this review and acknowledges interesting discussions 
with J. Ambj\o{}rn, J. Jurkiewicz, S. Nishigaki and J. Zinn-Justin. 


\smallskip

%
%

\end{document}